\documentclass[aps,prd,preprint,groupedaddress]{revtex4-1}
\usepackage{physics}
\usepackage{amssymb}
\usepackage{stix}
\def\hf{{\frac{1}{2}}}

\begin{document}

%\begin{frontmatter}

%\preprint{}

\title{Chaotic deterministic quantization in a 5D general relativity}


\author{Timothy D. Andersen}
\email{andert@gatech.edu}
%\thanks{}
\address{Georgia Institute of Technology, Atlanta, Georgia, 30332, USA}



\date{\today}

\begin{abstract}
How to quantize gravity is a major outstanding open question in quantum physics. While many approaches assume Einstein's theory is an effective low-energy theory, another possibility is that standard methods of quantization are the problem. In this paper, I analyze a quantization mechanism based on chaotic dynamics of 5D general relativity (with imaginary time) with BKL dynamics in the mixmaster universe as an example. I propose that the randomness of quantum mechanics as well as its other properties such as nonlocality derive from chaotic flow of 4D spacetime through a 5th dimension, with the metric tensor under Wick rotation to Euclidean space acting as a heat bath for other quantum fields. This is done by showing that the theory meets mixing conditions such that it is chaotically self-quantizing and quantizes other fields to which it is coupled, such that in the limit taking chaotic dynamics scale to zero the quantization is equivalent to a stochastic quantization. A classical stability analysis shows this dimension is likely spacelike. 
\end{abstract}

%\keywords{}


\maketitle

\section{Introduction}
In classical statistical mechanics, the source of randomness is clear. For example, Brownian motion of a grain of pollen can be described with a random forcing function, which appears in the original Langevin equation, but that forcing function derives from the random influences of tiny molecules. The motion of those molecules is apparently random but is in fact deterministic, described by nonlinear potential wells colliding in a fluid or gas. The convergence from chaotic determinism to stochastic equation is a property of rapid chaotic kick forces combined with a random initial condition \cite{Beck1990}.

The idea of using a 5th dimension to stochastically quantize fields in, e.g. a Langevin formalism using the Fokker-Planck equations has been around since the 1950s, with significant progress appearing in the 1980s \cite{Klauder1983}\cite{Namiki2008} with the Parisi-Wu mechanism being the simplest and most widely used \cite{Parisi1981}. In the early '80's Callaway introduced the chaotic or Hamiltonian quantization mechanism \cite{Callaway1982}\cite{Callaway1983}, recognizing the similarity to our understanding of Brownian motion. Both these methods were particularly well suited to fermionic lattice simulations.

Callaway's method had a major drawback versus the stochastic method, however, in that many quantum Hamiltonians did not describe ergodic systems. Hence, they did not cover their entire configuration space and ensure a correct quantization. Beck later developed a robust deterministic quantization based on Bernoulli shift or $\phi$-mixing kick forces which occur in nonlinear systems \cite{Beck2004}. But Beck's method only applies to fields coupled to certain kinds of scalar fields.

Extending this idea to all quantum fields, which are all similarly quantized with a temperature equivalent to $\hbar$ in imaginary time, requires first a universal medium, since all quantum fields and particles experience randomness. This medium must, therefore, be spacetime itself, since it is the only field that couples to every other field including itself. The second extension is the need to introduce a 5th dimension. This dimension may be spacelike or timelike but has a ``temporal'' nature in that spacetime flows in it based on an equilibrium flow.

In this paper, I propose a novel quantization mechanism based on a 5D general relativity with the 5D Einstein-Hilbert action in signature (-++++) and $x_4$ being the 5th dimension. The fifth dimension is not compactified, therefore, derivatives with respect to $x_4$ are non-zero and generate matter-like terms. I do not make the non-conformal Kaluza-Klein assumption of incorporating electromagnetism into the 5D metric, retaining, instead, higher dimensional matter. The matter-like terms do not interact with the electromagnetic field except gravitationally, indicating they are a form of ``dark'' matter and energy. 

Through a scaling analysis, I show that the nonlinearity of the 5D Einstein equations with a harmonic gauge condition increases with smaller length scales and vanishes at large scales, indicating that the metric becomes chaotic at small scales and has near linear wave-like solutions at largest scales. This may be sufficient to explain why gravity does not couple to vacuum fluctuations because it is the source of vacuum energy; thus there is a constant exchange between vacuum quantum fields and the gravitational, with the gravitational acting as the heat bath.

The paper is organized as follows: (1) I define the fifth dimension using coordinate free arguments that the fifth dimension is not a preferred direction but arises naturally out of Hamiltonian flow in a symplectic geometry. This Hamiltonian flow, for a 5D Hamiltonian, is similar to arguments defining time based on 4D Hamiltonians. (2) I establish the theory in terms of the 5D ADM Hamiltonian formalism (as opposed to the Kaluza-Klein approach which uses a non-conformal definition to derive electromagnetism). (3) I reformulate the theory in terms of the 5D harmonic coordinate gauge Einstein equations (a nonlinear wave equation) based on the Landau-Lifshitz form and apply renormalization scaling arguments to show how the equations become highly nonlinear at small scales and near-linear at large, and (4) I analyze the classical theory. A stability analysis of the spherically symmetric solution shows how the fifth dimension is likely spacelike. I also show how dark dust-like and vacuum-like terms emerge from higher dimensional derivatives, indicating possible sources of dark matter and energy in an isotropic-homogeneous solution.  (5) I conclude with a brief discussion of interpretation.

\section{Equilibrium Flow Definition of the Fifth Dimension}

Many quantization methods that rely on a 5th dimension add only a simple momentum-based term to the usual Langrangian, which they treat as a potential energy. This kinetic term can be integrated out of the path integral and, therefore, does not modify the original theory. For example, Callaway's Hamiltonian method adds a simple quadratic in derivatives with respect to the fifth dimension while Namiki's stochastic form uses a single derivative minus the original Lagrangian and squares the difference. While Namiki's approach is robust and widely applicable as it guarantees positive-definiteness, it does not apply well to general relativity where the new Lagrangian should be some recognizable scalar. The Parisi-Wu stochastic quantization is simpler and more applicable here but does not indicate the correct choice for the action. 

A natural choice is the 5D Einstein-Hilbert action which integrates the Ricci scalar over spacetime. This is closer to Callaway's approach but the kinetic term is the extrinsic curvature, which contains references to both the gravitational metric and its conjugate momenta. From the perspective of a symplectic geometry approach, this is the only choice to preserve the unique coordinate free aspects of general relativity. I will discuss how this changes the original theory in later sections.

While statistical mechanics typically assumes a pre-existing time dimension, Rovelli has shown that one can define a thermal time dimension based on a Hamiltonian flow using concepts of symplectic geometry defined with a two form $\omega = d\theta$ where $\theta$ is a one form defining general relativity. This thermal time is implicit in the flow of the Hamiltonian vector field rather than existing explicitly. In other words, the Gibbs state $\rho\propto e^{-\beta H}$ is defined independently of any pre-defined time dimension. Thus, rather than $H$, the Hamiltonian being dependent on time, physical equilibrium determines the form of H such that time appears as a flow. The following discussion of the 5th dimension follows along the lines of the generally covariant statistical mechanics found in \cite{Rovelli2013}.

Let ${\cal E}$ be a symplectic space and $C$ a submanifold upon which the dynamical system is defined. An ergodic dynamical system follows a microcanonical statistical distribution $\rho_{micro}\propto \delta(E-H)$ for some energy $E$. Let $\rho_{micro}$ be a normalized measure on the configuration space $\Gamma$, $\int_{\Gamma} \rho_{micro} = 1$. A Hamiltonian vector field $X$ is defined by $\omega(X) = dH = 0$ which for a choice of $E$ determining the initial condition generates a flow in phase space with a parameter $\tau$. This flow, however, is not unique since any statistical state may generate a flow. (This is reflected in that the ADM Hamiltonian is a general form of the Einstein equations with no special coordinate system.) The remaining ingredient to define a unique flow is for the system to be in equilibrium, which we can define as a maximally entropic or equilibrium sequence of states that satisfy detailed balance. This is, in discrete systems referred to as transition rate symmetry, i.e., the frequency of transitions from a state $i$ to a state $j$ is equal to the frequency of the reverse. In continuous systems, it means that the probability of a path $P$ made up of points in $\Gamma$ from a state $\rho$ to a state $\rho'$ is equal to the probability of the reverse path $P'$. This is sufficient to define a statistical theory and fix the fifth dimension to one that is along an equilibrium path.

The parameter $\tau$ is referred to as a local thermal time. And given a set of coordinates $\gamma,\phi$ and their conjugate momenta $\pi,p$ for the induced 4-D metric and matter fields respectively from a 4D locally Euclidean manifold $\Sigma$ to a target space $V$, an orbit $o$ is a solution to the field equations in the 5D pseudo-Riemannian manifold $(M,g)_o$. This orbit determines a submanifold $\Phi:\Sigma \rightarrow (M,g)_o$ with metric $\gamma$ and conjugate momenta $\pi$ determining extrinsic curvature. Then $\Phi_\tau$ is a foliation corresponding to the line of the orbit.

From this, one can choose a coordinate system for the 5-D Lagrangian under Wick rotation such that $x_4=\tau$ corresponds to an equilibrium flow satisfying detailed balance.

\section{Equations of Motion}
The 5-D formalism under Wick rotation is a 5-D metric, $g_{AB}$ where capital Latin letters will refer to 5-D indexes $A,B=0,1,2,3,4$, lower case Greek letters will refer to 4-D indexes $\alpha,\beta=0,1,2,3$, and 3-D spatial indexes will be lower case Latin letters $i,j=1,2,3$. The ADM form we will use is
\begin{eqnarray}
	g_{44} & = & -\epsilon\phi^2 + \gamma^{\alpha\beta}A_\alpha A_\beta,\\
	g_{4\alpha} & = & A_\alpha,\quad g_{\alpha\beta} = \gamma_{\alpha\beta},
\end{eqnarray} with inverses $g^{44} = -\epsilon/\phi^2$, $g^{4\alpha} = A^\alpha/\phi^2$, and $g^{\alpha\beta} = \gamma^{\alpha\beta} - \epsilon A^\alpha A^\beta/\phi^2$. The signature of $g_{44}$ is given by $-\epsilon$ with $\epsilon=1$ timelike and $\epsilon=-1$ spacelike.

Covariant derivatives are with respect to the submanifold metric $\gamma_{\alpha\beta}$, e.g., $\nabla_\alpha A^\beta = \partial_\alpha A^\beta + \gamma^\beta{}_{\beta\mu} A^\mu$ where $\gamma^\beta{}_{\beta\mu} = \hf \gamma^{\alpha\nu}(\partial_\beta \gamma_{\mu\nu} + \partial_\mu \gamma_{\beta\nu} - \partial_\nu \gamma_{\beta\mu})$ is the Christoffel symbol on the submanifold.

The extrinsic curvature is defined as $K_{\alpha\beta} = \frac{1}{2\phi}(\nabla_\alpha A_\beta + \nabla_\beta A_\alpha - \partial_4 \gamma_{\alpha\beta})$.

The ADM Lagrangian is given by,
\begin{equation}
	{\cal L}_{ADM} = \phi\sqrt{\gamma}(\epsilon (K_{\alpha\beta} K^{\alpha\beta} - K^2) + {}^{(4)} R),
\end{equation} where $K=\gamma^{\alpha\beta} K_{\alpha\beta}$ is the trace and ${}^{(4)} R$ is the 4D Ricci Scalar. 

In the ADM formalism we typically go on to define the Hamiltonian from here, but for path integral quantization this Lagrangian is our Hamiltonian and the ADM Hamiltonian is only useful if we were to go on to a canonical operator formalism with Ashtekar variables.

Instead, we need to eliminate the second derivatives with respect to $x_4$ and redefine the extrinsic curvature in terms of conjugate momenta, $\pi^{\alpha\beta} = \sqrt{\gamma}(K\gamma^{\alpha\beta} - K^{\alpha\beta}) = \sqrt{\gamma} P^{\alpha\beta}$,
\begin{equation}
	{\cal L}_{ADM} = \frac{\epsilon\phi}{\sqrt{\gamma}}(\pi^{\alpha\beta} \pi_{\alpha\beta} - \hf \pi^2) + \phi\sqrt{\gamma}\,{}^{(4)}R,
\end{equation} where $\pi = \gamma^{\alpha\beta} \pi_{\alpha\beta}$.

With units such that $\hbar=1$, the path integral may now be defined as (suppressing indices),
\[
Z = \int D\{\pi,\gamma,p,\psi\} e^{-S_{ADM} - S_M},
\] where $S_{ADM} = \int d^5x {\cal L}_{ADM}$ is the action and $S_M = \int d^5x {\mathfrak L}_M$ is the matter action for some matter fields $\psi$ with Lagrangian ${\cal L}_M$ and $p=\partial_4 \psi$.

This path integral is equivalent to the microcanonical quantization $Z\propto\Omega$ with $\Omega = \int D\{\pi,\gamma,p,\psi\} \delta(A - S)$ for $S=S_{ADM} + S_M$ with an appropriate regularization and choice of fixed action $A$ \cite{Strominger:1983}. For ergodic systems this is equivalent to the average over infinite $x_4$ of the solution to the equations of motion generated by $S_{ADM}$ and $S_M$,
\begin{equation}
	\partial_4 \gamma = \{\gamma, S\},\quad \partial_4 \pi = \{\pi, S\},\quad\partial_4 \psi = \{\psi, S\},\quad\partial_4 p = \{p, S\}
	\label{eqn:pbra}
\end{equation} for suitable initial condition as to guarantee a temperature of $\hbar$. (One can also explicitly add a stochastic term as in stochastic quantization and obtain the standard Langevin quantization \cite{Klauder1983}\cite{Namiki2008} but this does not suit the purpose of this paper to show a deterministic quantization scheme.)

The Poisson bracket is defined to be,
\begin{equation}
\{A,B\} = \int d^4x \frac{\delta A}{\delta \gamma_{\alpha\beta}}\frac{\delta B}{\delta \pi^{\alpha\beta}} - \frac{\delta A}{\delta \pi^{\alpha\beta}}\frac{\delta B}{\delta \gamma_{\alpha\beta}}.
\label{eqn:poisson}
\end{equation} The variations are not standard variations as when we derive the Einstein equations because now $\pi^{\alpha\beta}$ is an independent degree of freedom. Thus, we will get equations that are different from the 5D Einstein field equations.

Deriving the equations of motion from the brackets, we have that
\begin{eqnarray}
	\epsilon\partial_4 \pi^{\alpha\beta} & = & \phi\sqrt{\gamma}[{}^{(4)}R^{\alpha\beta} - \hf {}^{(4)}R\gamma^{\alpha\beta}] -  
	(\nabla^\alpha\nabla^\beta - \gamma^{\alpha\beta}\nabla^2)\phi  + \nonumber\\& & \epsilon\frac{\phi}{\sqrt{\gamma}}\left[2\pi^{\alpha}_\mu \pi^{\beta\mu} - \pi \pi^{\alpha\beta} - \hf\left(\pi_{\alpha\beta}\pi^{\alpha\beta} - \hf \pi^2\right)\gamma^{\alpha\beta}\right] - \frac{\delta S_M}{\delta\gamma^{\alpha\beta}}
	\label{eqn:pi}
\end{eqnarray} and
\[
\partial_4\gamma_{\alpha\beta} = - \frac{\phi}{\sqrt{\gamma}}(2\pi_{\alpha\beta} - \pi\gamma_{\alpha\beta}).
\] If we let $\partial_4\cdot = 0$, $\phi=1$, and $A^{\alpha}=0$, we recover the 4D Einstein equations.

There are a few differences between these equations of motion and the standard 5D Einstein equations or their ADM equivalent. First, they are missing a few dependencies on $A^{\alpha}$ that would ordinarily come out of the Legendre transform. The only dependencies are inside of $\pi^{\alpha\beta}$. The other difference is that from the standard equations, the sign of the $\partial^4\partial_4 \gamma_{\alpha\beta}$ term is reversed. This means that a spacelike fifth dimension will lead to a 4+1-D wave equation while a timelike one will lead to a 5-D Poisson equation in the linearized equations. Since the signs of the other terms are unchanged from their standard form, however, other implications of the signature are also unchanged including the sign of the extrinsic curvature.

The reason why the sign is different is because, in a stochastic or chaotic quantization, the kinetic energy represents the stochastic variations away from the classical Lagrangian. Thus, increases in kinetic energy lead to increases in the classical Lagrangian with zero kinetic energy representing the classical least action. In a typical Hamiltonian, however, increases in kinetic energy lead to decreases, not increases, in potential energy.

\section{Chaotic Quantization}

A strongly self-interacting field in five dimensions under certain mixing assumptions generates fast time chaotic dynamics at $\tau$ scale $\sim\delta$ that become equivalent to a white noise Weiner process as $\delta\rightarrow 0$. It has been rigorously proved that chaotic quantization under these mixing assumptions is equivalent to standard quantization via equivalence to Parsi-Wu stochastic quantization \cite{Beck1995}.

In this section I show that the 5D general relativistic theory so far presented satisfies these mixing assumptions. This will imply that the theory is both self-quantizing and quantizes all other fields to which it is coupled.

The primary hypothesis for the quantum theory is that, in imaginary time, chaotic dynamics causes the gravitational field to act as a heat bath for other quantum fields. This implies that vacuum energy is the result of energy exchanges between the gravitational field and other quantum fields occurring at small scales. I conjecture that these effects are not observed cosmologically because most vacuum energy effects vanish at larger scales, and there is no cumulative gravitational effect because small scale exchanges screen them out. 

\subsection{Asymptotic Scaling and Self-Interaction of Gravitational Waves}
The difference between the quantum and classical versions of the theory is one of scale. While a rigorous treatment of renormalization group flow is beyond the scope of this paper, I will show scaling is possible using a Landau-Lifshitz formalism which allows the equations \ref{eqn:pi2} to be converted into a nonlinear wave equation. This scaling demonstrates strong self-interaction at small scales required for self-quantization as well as quantization of the other forces and matter.

The Einstein-Hilbert action is not perturbatively renormalizable, in that, in 4D, one cannot, with just a few counter-terms, eliminate the energy cutoff in the perturbation series. Kupiainen refers to this condition as superrenormalizability \cite{Kupiainen2016} to emphasize that it is much stronger than what is needed to renormalize a theory. While it is a condition of quantum fields of the Standard Model, stochastic and dynamical (PDE-based) systems that are not superrenormalizable are often still renormalizable with the appropriate choice of scaling \cite{Gawdzki1985}. While a turnkey methodology of rigorous renormalization has so far eluded mathematicians, such group flows have been proved for nonlinear parabolic PDEs both stochastic \cite{Kupiainen2016} and non-stochastic, e.g., the Ginzburg-Landau equation, \cite{Bricmont1992}\cite{Bricmont1994} as well as the stochastic wave equation in 2D \cite{Gubinelli2018}. And considerable progress has been made in the last 30 years in this domain \cite{Giuliani2021}.

Wilson's formulation of renormalization for non-perturbative systems is a general approach that relies on iterative scaling of the system and developing either a recurrence relation between the system at one scale $L^n$ and the next scale $L^{n+1}$ or a set of differential equations in the case of a continuous scaling function with infinitesimal changes in scale $L(1+\delta)$. This is for some scale factor (typically $L<1$ for the discrete version) such that $L^N$ is the cutoff scale \cite{Kupiainen2016}\cite{Wilson1983}. While linear systems tend to be scale invariant, nonlinear systems introduce scaling effects that require renormalization of their coupling constants. A system can be non-chaotic at one scale while chaotic at another with some transition from one to the other at some scale and this is reflected by increasing or decreasing coupling strengths dialing up or down nonlinear effects. In the following, I show that nonlinear effects for gravity increase at smaller scales and that this causes a transition from largely smooth, wave-like behavior at large scales to chaos at small. This explains why quantum effects are typically observed at small scales only as in Brownian motion. It is the typical scale at which the gravitational field is highly nonlinear that determines the scale at which quantum effects are observed in our universe.

An important goal of renormalization is to extend the equation \ref{eqn:pi2} to infinite $\tau$ and describe its behavior. I will not prove but assume that with appropriate spatial boundary conditions, e.g., asymptotic flatness or homogeneity, and an initial condition at $\tau=0$ the equations have a regular solution for finite $\tau=L$. Quantization demands that averaging be done over infinite $\tau$, $\langle O\rangle = \lim_{L\rightarrow\infty} \frac{1}{L}\int_0^L d\tau O[\bar{\gamma},\bar{\pi}]$ for some observable functional, $O$, of a solution to \ref{eqn:pi}, $\bar{\gamma}$ and  $\bar{\pi}$ (with indexes suppressed for brevity). As I showed in previous work \cite{Andersen2019}, vacuum loops that cause blow-ups in quantization do not occur when 5D PDEs that have regular solutions are solved for or averaged over finite $\tau$ but, rather, appear in taking the limit to infinite $\tau$. Thus, a cutoff $L$ for $\tau$ serves as a regularization in this scheme in the same way that a length, energy, or dimensional regularization can be used in a path integral. In this case, the solution will be cutoff dependent and the goal is to remove the dependence through a renormalization of coupling constants.

We have two running coupling constants. The first is the gravitational coupling $\kappa = 8\pi G/c^4$ which I have taken to be $1/2$ in the previous sections but will introduce explicitly in this section. The second is the gravitational self-coupling. In order to define this coupling, we must first put the equations \ref{eqn:pi2} into what is sometimes called ``relaxed'' form \cite{Will2014}. Let $h^{\alpha\beta} = \eta^{\alpha\beta} - (\gamma)^{1/2} \gamma^{\alpha\beta}$ represent deviations from flatness scaled by the volume element. We assume that the region of interest can be covered by harmonic coordinates. Choose a harmonic gauge condition $\partial h^{\alpha\beta}/\partial x^\beta = 0$, similar to the Lorentz gauge for Maxwell's equations. Assume time is imaginary and spacelike. The equations \ref{eqn:pi2} become,
\begin{equation}
	\square_5 h^{\alpha\beta} = -2\kappa\gamma T^{\alpha\beta} - 2\epsilon {\mathfrak u}^{\alpha\beta} - {\mathfrak t}^{\alpha\beta},
\end{equation} where $\square_5 = -\partial^2/\partial\tau^2 + \nabla_4^2$, $\nabla_4 = \partial/\partial x^\mu$, and ${\mathfrak t}^{\alpha\beta}$ is the non-linear (intrinsic) contribution of the gravitational field given by the scaled Landau-Lifshitz pseudotensor, $\tau^{\alpha\beta}_{L-L}$ (\cite{Misner1973}, eqns. 20.20 and 20.21), ${\mathfrak t}^{\alpha\beta} = 16\pi\tau^{\alpha\beta}_{L-L}/(\gamma)$. The extrinsic curvature contribution is ${\mathfrak u}^{\alpha\beta}$. The L-L pseudotensor is made up of terms that are quadratic in ordinary derivatives of $h^{\alpha\beta}$, e.g., $h^{\alpha\beta}{}_{,\lambda} h^{\lambda\mu}{}_{,\mu}$ and instances of $\gamma^{\alpha\beta}$ and $\gamma_{\alpha\beta}$ that are paired with one another \cite{Misner1973}. This means that any scaling of $h^{\alpha\beta}$ (as well as corresponding scaling of $\gamma^{\alpha\beta}$ and $\gamma$) and coordinates $x^\mu,\tau$ will scale the entire pseudotensor by a power of the scale factor. It turns out that ${\mathfrak u}^{\alpha\beta}$ scales the same way provided one reintroduces $\phi$ from \ref{eqn:pi} as a constant and scales it by the same factor. (The momentum $\pi^{\alpha\beta}$ scales like $h^{\alpha\beta}$.) 

The choice of scaling will determine the power. For example, a scaling of $h^{\alpha\beta}(x,\tau)\rightarrow \lambda h^{\alpha\beta}(x,\tau)$ with $\lambda>0$ results in
\begin{equation}
	\lambda\square_5 h^{\alpha\beta} = -2\lambda\kappa'\gamma T^{\alpha\beta} - 2\epsilon\lambda^2 {\mathfrak u}^{\alpha\beta} - \lambda^2 {\mathfrak t}^{\alpha\beta},
\end{equation} where $\kappa'$ is rescaled. Dividing by $\lambda$,  there is a single self-coupling constant. Thus, the equations become
\begin{equation}
	\square_5 h^{\alpha\beta} = -2\kappa\gamma T^{\alpha\beta} - \lambda(2\epsilon {\mathfrak u}^{\alpha\beta} + {\mathfrak t}^{\alpha\beta}),
	\label{eqn:wave}
\end{equation} where we drop the prime from $\kappa'$ now. These equations are solved by the integral equation,
\begin{equation}
	h^{\alpha\beta} = G\left[-2\kappa\gamma T^{\alpha\beta} - \lambda(2\epsilon {\mathfrak u}^{\alpha\beta} + {\mathfrak t}^{\alpha\beta})\right] + G_0\left[h_0^{\alpha\beta}\right],
\end{equation} where $G$ is the Green's function operator for the $4+1$-D wave equation,
\begin{equation}
	(Gf)(x,\tau) = \frac{1}{8}\int_0^\tau d\sigma \intbar_{B(x,\tau-\sigma)} dy\,f(y,\sigma)G(x-y,\tau-\sigma),
\end{equation} and $G_0$ is the functional for the initial condition $h_0^{\alpha\beta} = h^{\alpha\beta}(x,0)$ and $\partial h^{\alpha\beta}/\partial \tau = 0$ at $\tau=0$. The Green's function has the form
\[
G(x,\tau) = 4\tau D(x,\tau) - \tau D^3(x,\tau).
\] such that $D(x,\tau) = \frac{\tau}{(\tau^2 - |x|^2)^{1/2}}$ and the initial condition functional has the form,
\[
(G_0f_0)(x,\tau) = \intbar_{B(x,\tau)} dy\,f_0(y)[D(x-y,\tau) - D^3(x-y,\tau) + D^5(x-y,\tau)/8].
\]

As in Callaway's computational approach, the initial condition ensures that the temperature of the statistical ensemble is at $\hbar$ by fixing the action of the microcanonical ensemble \cite{Strominger:1983}\cite{Callaway1983}. It is typical to choose an initial condition from a statistical distribution but any specific details of the initial condition will be ``forgotten'' asymptotically because of mixing. Therefore, any will do, and it is simplest to place all the action in the ground state, $k^\mu = 0$, such that $h_0^{\alpha\beta}(x_\mu) = \int \frac{d^4k}{(2\pi)^4} \hat{C}_0^{\alpha\beta}\delta(k^\mu) e^{-ik^\mu x_\mu} = \hat{C}_0^{\alpha\beta}$ meaning that it is constant in spacetime. Another option is to place it all at a single point, $h_0^{\alpha\beta}(x_\mu) = C_0^{\alpha\beta} \delta(x_\mu)$ so it becomes constant in $k_\mu$.

Let $h_L^{\alpha\beta}(x,\tau) = L^p h^{\alpha\beta}(Lx,L\tau)$, for some $p$, be a rescaling of the field. Given equation \ref{eqn:wave}, $h_L^{\alpha\beta}$ satisfies,
\begin{equation}
	\square_5 h_L^{\alpha\beta} = -2\kappa\gamma L^{p + 2} T^{\alpha\beta} - \lambda L^{-p}(2\epsilon {\mathfrak u}_L^{\alpha\beta} + {\mathfrak t}_L^{\alpha\beta}),
	\label{eqn:waveL}
\end{equation} Thus, the matter coupling scales like $\kappa_L = \kappa L^{p+2}$ and $\lambda_L=\lambda L^{-p}$. Since $\gamma$ implicitly depends on $h^{\alpha\beta}$ from its definition, it also depends implicitly on $\lambda$. Solving for $\gamma$ from the definition of $h$, $\gamma = [(\eta^{\alpha\beta} - \lambda h^{\alpha\beta})\gamma_{\alpha\beta}/4]^2$. Thus, $\gamma = [(\eta^{\alpha\beta} - \lambda_L L^p h^{\alpha\beta})\gamma_{\alpha\beta}/4]^2 = [(\eta^{\alpha\beta} - \lambda_L h_L^{\alpha\beta})\gamma_{\alpha\beta}/4]^2$.

In the language of renormalization, a Renormalization Group (RG) map $R_L$ on the set $(f^{\alpha\beta},\lambda,\kappa)$ where, for convenience in scaling $\tau$ $f^{\alpha\beta}(x) = h^{\alpha\beta}(x,1)$ is the initial condition. The map has the following relationship
\[
R_L(f^{\alpha\beta},\lambda,\kappa) = (h_L^{\alpha\beta}(x,1),\lambda L^{-p}, \kappa L^{p+2})
\]

The goal of the map is to show how scaling of the initial condition and couplings produces a solution which itself can become the new initial condition and so on, approaching an asymptotic fixed point.

Suppose $\lambda=0$. Then $\gamma\rightarrow 1$ and we have a linear, inhomogenous wave equation
\begin{equation}
	\square_5 h_L^{\alpha\beta} = -2\kappa_L T^{\alpha\beta}.
	\label{eqn:waveLaymp}
\end{equation} This suggests that renormalization can lead to simple, linear behavior at large scales which indicates a reasonable limit.

At small scales, however, it is clear that the field is more strongly self-interacting while more weakly interacting with matter.

\subsection{Dynamical system of nonlinear Langevin type}
The one-dimensional dynamical system of nonlinear Langevin type takes the form,
\begin{equation}
	\dot{Y} = A(Y) + L_\delta(\tau),
	\label{eqn:langevin}
\end{equation} for a system $Y(\tau)$ and a nonlinear evolution equation $\dot{Y} = A(Y)$.

The function $L_\delta(\tau)$ provides a deterministic but chaotic kick force where
\begin{eqnarray}
	L_\delta(\tau) & = & \delta^{1/2}\sum_{n=1}^\infty x_n\delta(t - n\delta),\\
	x_{n+1} & = & T(x_n)
	\label{eqn:kick}
\end{eqnarray} The kick strengths $x_n$ evolve in a deterministic way under the map $T$. The dynamical system \ref{eqn:langevin} converges to a stochastic Langevin equation as $\delta\rightarrow 0$ provided the initial kick $x_0$ is distributed according to a smooth probability distribution and the map $T$ has a $\phi$-mixing property which is a feature of Bernoulli shift mixing \cite{Beck1990}. Second and third order Chebyshev polynomials ($2x^2-1$ and $4x^3-3x$) are examples that satisfy this property.

In the case of the vacuum solutions, the self-coupling parameter $\lambda$ scales the nonlinearity and as it becomes larger one expects there to be a transition from smooth gravitational wave interactions to turbulence. Gravitational wave turbulence in weak nonlinearity has been studied in the context of the early universe \cite{Galtier2017}.

The dominant resonant interactions for weak gravitational waves is four wave or higher, which requires expansion to third order in the nonlinearity. Beyond weak to strong interactions where $\lambda > 1$, for example, weak nonlinear wave expansions are no longer appropriate and a chaotic mapping approach to an exact solution is required as in the Mixmaster universe.

In the context of General Relativity, the full equations prove intractable for a complete study of mixing and chaos. Many simplified models have been shown not to be chaotic while a few have been but perturbations may introduce chaos into non-chaotic models \cite{Misner1969}\cite{Barrow1982}. 

The Mixmaster universe, a homogeneous, anisotropic solution to the Einstein equations discovered in the 1960s, has been studied extensively because it was not clear if its chaotic behavior was a real pheonoemnon or observer dependent. Its chaos was eventually shown to be observer independent in the 1990s \cite{Cornish1997}. Therefore, it is a good choice to demonstrate how chaotic behavior can lead to Langevin-type dynamics.

Suppose we have a free scalar matter field $\psi$ given by the Klein-Gordon Lagrangian density,
\begin{equation}
	{\cal L}_{KG} = \sqrt{-g}\hf \left( g^{AB}\partial_A \psi \partial_B \psi - m^2\psi^2\right).
\end{equation} 

The momentum is defined as,
\begin{equation}
	\Pi = \frac{\partial{\cal L}_{KG}}{\partial (\partial_4 \psi)} = \sqrt{-g}\left( \partial^4\psi + g^{\mu 4}\partial_\mu\psi\right).
\end{equation}
In order to quantize this with a deterministic, chaotic Langevin-type approach, we must replace instances of $\sqrt{-g}\partial^4\psi$ with $\Pi$ and define the action, $S_{KG} \equiv \int d^4x\, {\cal L}_{KG}(\Pi,\psi)$. We can assume $g^{44} = 1$ since this only corresponds to coordinates that stretch and compress with the manifold in the $x^4$ dimension.
\[
S_{KG} = \int d^4x \hf \Pi\partial_4\psi + \sqrt{-g}\hf \left( \gamma^{\mu\nu}\partial_\mu \psi \partial_\nu \psi - m^2\psi^2\right).
\] or
\[
S_{KG} = \int d^4x \hf \frac{\Pi^2}{\sqrt{-g}} - \hf \Pi g^{\mu 4}\partial_\mu\psi + \sqrt{-g}\hf \left( \gamma^{\mu\nu}\partial_\mu \psi \partial_\nu \psi - m^2\psi^2\right).
\] This action then gives the equations of motion in the 5th dimension by Hamilton's equations, $\partial_4\Pi = \{\Pi,S_{KG}\}$ and $\partial_4\psi = \{\psi,S_{KG}\}$, and an appropriate initial condition fixes the action such that the temperature is, on average, $\hbar$.

Now we assume ``synchronous'' coordinates where the synchronization is in the fifth dimension $\tau$ and not time. Therefore, $g^{\mu 4} = g^{4\mu} = 0$ and $\sqrt{-g} = \sqrt{-\gamma}$. Now, the equations of motion are,
\begin{equation}
	\partial_4\Pi = -\frac{\delta V}{\delta \psi}
\end{equation} where $V=\sqrt{-g}\hf \left( \gamma^{\mu\nu}\partial_\mu \psi \partial_\nu \psi - m^2\psi^2\right)$ is the standard Klein-Gordon Lagrangian in four dimensional geometry and $\partial_4\psi = \Pi/\sqrt{-\gamma}$.

Since $\frac{\delta V}{\delta \psi}$ is simply the 4-D Klein-Gordon equations we have,
\begin{equation}
	\partial_4\Pi = (\square - m^2)\psi
\end{equation} where $\square = \gamma^{\mu\nu}\nabla_\mu\partial_\nu$ and $\nabla_\mu$ is the 4-D covariant derivative, and, recalling that we are in Euclidean space, the signature is $(++++)$. Therefore, as in the equations for quantizing general relativity, the spacelike fifth dimension behaves like a temporal dimension in switching sign.

The covariant derivative means that the Klein-Gordon equation has the form,
\[
\partial_4\Pi = \gamma^{\mu\nu}\partial_\mu\partial_\nu\psi - \gamma^{\mu\nu}\Gamma^\lambda{}_{\mu\nu}\partial_\lambda\psi - m^2\psi.
\] In a chaotic geometry, let $\gamma_{\mu\nu} = \eta_{\mu\nu} + \xi_{\mu\nu}$ where $\eta$ is the flat background geometry and $\xi$ is the remaining, chaotic part. The additive noise is given by the chaotic contribution,
\[
L(\tau,x) = \xi^{\mu\nu}\partial_\mu\partial_\nu\psi - \gamma^{\mu\nu}\Gamma^\lambda{}_{\mu\nu}\partial_\lambda\psi,
\] 
(Since they do not contain instances of $\Pi$ or $\partial_4$ terms, they are not multiplicative noise.) The drift of the Langevin-type equation is given by the flat geometry equation $(\eta^{\mu\nu}\partial_\mu\partial_\nu-m^2)\psi$.

While a general demonstration of quantization from general relativity is not possible, it can be demonstrated from a chaotic Belinskii, Khalatnikov and Lifshitz (BKL) oscillation of the mixmaster type close to singularities which occurs in pure gravity formulations up to 11 dimensions \cite{Damour2001}. This result indicates that if dynamics in the fifth dimension cause these kinds of singularities at small spacetime scales, then chaotic oscillations result.

Barrow showed the mixmaster model to be a Bernoulli system \cite{Barrow1982}. The BKL dynamics have positive-metric and topologic-entropy, weak Bernoulli properties, and are strongly mixed and ergodic \cite{Montani2008}.

The main difference between this and standard formulations is that we swap the time and fifth dimension, $\tau$, so that the chaotic behavior occurs relative to $\tau$, which has a temporal nature, while the imaginary time dimension, being spacelike, has a scale factor that participates in the chaotic oscillations like those of the other three spatial dimensions.

We can show the quantization from these chaotic dynamics in the case of the free scalar Klein-Gordon if we discretize the equation,
\begin{equation}
	\partial_4\Pi = (\eta^{\mu\nu}\partial_\mu\partial_\nu-m^2)\psi + L(\tau,x),
\end{equation} with a grid spacing of $\delta_A$ and defining $\tau=n\delta_4$,
\begin{eqnarray}
	\frac{\Pi((n+1)\delta_4) - \Pi(n\delta_4)}{\delta_4} & = & \eta^{\mu\nu}\frac{\psi(x + \delta_\mu + \delta_\nu) - \psi(x + \delta_\mu) - \psi(x + \delta_\nu) - \psi(x)}{\delta_\mu\delta_\nu} -\nonumber\\ & & m^2\psi(x) + L_\delta(\tau,x).
	\label{eqn:discrete}
\end{eqnarray}

Given that $L_\delta$ is defined by the discrete BKL approximation to the Mixmaster dynamics in the form given by \ref{eqn:kick} and a fixed choice of initial condition $\phi_0$, the system is fully deterministic. If the initial condition is selected from a probability distribution, then $\phi$ is a stochastic process. For finite $\delta_A$, that process is complex and non-Gaussian with higher order correlation functions.

A simplification occurs, however, when we take $\delta_A\rightarrow 0$, and the process converges to an Ornstein-Uhlenback process if $T$ in \ref{eqn:kick} is a Bernoulli system (i.e., a system with the randomness property of a Bernoulli shift which is a common feature of pseudorandomn number generators) \cite{Beck1990}\cite{Beck2004}. Beck uses this feature to quantize fields with a scalar field with Bernoulli dynamics, but any field with this feature will do in the limit $\delta_A\rightarrow 0$. As Beck shows, this simplification enables equations of the form \ref{eqn:discrete} to converge to the correct quantization in the limit.

For other fields including vector fields, the procedure is the same: derive evolution equations that describe the dynamics of the 4D system in the fifth dimension and demonstrate that chaotic dynamics in the metric field are sufficient to quantize the system by converging to a Parisi-Wu stochastic quantization. The path integral formulation then becomes an approximation of the statistics.

\section{Classical Theory}
We show that the classical theory is in agreement with observation. The equations of motion \ref{eqn:pi} are identical to the Einstein field equations in 4D for $\partial_4\cdot = 0$, $\phi=1$, and $A^{\alpha}=0$. If these are within some small $\delta$ of those values, then, all observations that agree with Einstein's will be met. Quantum theory, we will see, requires $\partial_4$ to be large at small scales and we will deal with that via renormalization in a later section but at large scales it must be relatively small.

The first point to note is that we can set $A^\alpha=0$ without loss of generality since it does not appear in the equations directly. This is a consequence of the coordinate freedom within the statistical distribution. As seen in the previous section, the statistical distribution only depends on $\pi^{\alpha\beta}$,$\gamma^{\alpha\beta}$, and $\phi$.  The only reason we would need to reintroduce $A^{\alpha}$ is if we solved for $\gamma^{\alpha\beta}$ and then later changed coordinate systems of the 5D manifold metric that resulted in the off-diagonals becoming non-zero. This means that the geodesic equation for a test particle is the same as the 4D geodesic, $\partial^2 x^\alpha/\partial\sigma^2 = - \gamma^\alpha{}_{\beta\mu}(\partial x^\beta/\partial\sigma)(\partial x^\mu/\partial\sigma)$.

In addition, we can now assume extrinsic curvature derives from $x_4$ variations in the 4D metric: $K_{\alpha\beta} = -\partial_4\gamma_{\alpha\beta}/(2\phi)$. These variations generate mass-like terms similar to how time variations in the metric in standard GR create mass-like terms in Newtonian gravity, e.g., from the energy of gravitational waves or energy from the expansion of the universe in cosmology. In Newtonian gravity, these are typically small enough to be neglected (unless we are looking at black hole mergers or rapidly orbiting pulsars). In this case, the masslike terms are a result of quantum deviations from the classical action that all matter and fields in the universe experiences.

We can also show that we can choose a coordinate system where $\phi^2 = 1$. Consider a family ${\cal O}$ of observables  O on $\Gamma$. The mean value of $O$ on a state $\rho$ is, $\langle O\rangle = \int_\Gamma O\rho$. The change in $O$ is given by $\partial_4 O = \{O,S\}$. Define the mean geometry $\bar{g}$ of a state $\rho$ for an observable as a spacetime $(M,\bar{g})$ with foliation $\phi_\tau$ such that $\langle O\rangle(\tau) = O(\phi^{-1}_\tau(\bar{g}))$, meaning that this mean geometry generates expected values of observables. It follows that $(M,\bar{g})$ is stationary under the flow defined by $\phi_\tau$. Therefore $\xi = \partial/\partial\tau$ is a Killing field that is timelike for a timelike $x_4$ or spacelike for a spacelike $x_4$ on $(M,\bar{g})$. 

This argument is typically applied to time and world lines (e.g., geodesics) but applies equally well here \cite{Rovelli2013}. Under Wick rotation, with time having spacelike dimension, a local coordinate system on, say, a point on a particle worldline does not flow in time but only in the dimension defined by statistical equilibrium. This point has a path $P$ defined by the equilibrium flow and has its own Euclidean coordinate system defined by a direction along the line and directions perpendicular to it (with the freedom to define 4D rotations about it). The coordinate parameter defining this proper dimension (I avoid calling it ``time'') $s$ has the following relationship to both the thermal flow paramter $\tau$ and the Killing field: The norm of $\xi$ is the ratio of the local proper dimension $s$ and the thermal flow $\tau$. This means that, from the Tolman-Ehrenfest law, $\phi^2 |\xi|$ is constant in stationary coordinates, $(x^\alpha,\tau)$, since $ds^2 = \phi^2 d\tau^2$. Thus, we can take $\phi^2 = 1/\hbar = 1$ in these coordinates. 

With this assumption, references to $\phi$ drop out of the equations. Thus, we have shown that of the three deviations from standard gravitational theory, only the depedencies on $\partial_4\gamma_{\alpha\beta}$ remain and the extrinsic curvature is exactly $K_{\alpha\beta} = -\partial_4\gamma_{\alpha\beta}/2$. This means that any Brans-Dicke theory predictions are ruled out completely which is in good agreement with the most recent observations \cite{Amirhashchi2020}.

From this we can conclude that the only difference between standard theory and this theory is that we should observe matter or energy (which can be either radiation-like, dust-like, or a mixture of the two) that has gravitational effects but does not interact with the electromagnetic field except gravitationally.

The equations \ref{eqn:pi} with $\phi=1$ can be written as:
\begin{equation}
	 {}^{(4)}R^{\alpha\beta} - \hf {}^{(4)}R\gamma^{\alpha\beta}  = \hf T^{\alpha\beta} + \epsilon U^{\alpha\beta}
	\label{eqn:pi2}
\end{equation} where
\[
U^{\alpha\beta} = \frac{1}{\sqrt{\gamma}}\partial_4 \pi^{\alpha\beta} - \frac{1}{\gamma}\left[2\pi^{\alpha}_\mu \pi^{\beta\mu} - \pi \pi^{\alpha\beta} - \hf\left(\pi_{\mu\nu}\pi^{\mu\nu} - \hf \pi^2\right)\gamma^{\alpha\beta}\right]
\] and $\delta S_M/\delta \gamma_{\alpha\beta} = \hf \sqrt{\gamma}T^{\alpha\beta}$. Writing $\pi^{\alpha\beta}$ in terms of $\gamma_{\alpha\beta}$:
\[
\pi^{\alpha\beta} = \sqrt{\gamma}(\gamma^{\alpha\mu}\gamma^{\beta\nu}\partial_4\gamma_{\mu\nu} -\gamma^{\alpha\beta}\gamma^{\mu\nu}\partial_4 \gamma_{\mu\nu}).
\]

\subsection{Static Spherically Symmetric}
There is a considerable body of research on the spherically symmetric Kaluza-Klein equations in both the compactified and non-compatified approaches. For a review see \cite{Overduin:1997}. Moreover, stability of spherically symmetric solutions of black holes and branes in arbitrary dimensions has been studied as well \cite{Prabhu2015} with general stability conditions given in terms of canonical energy. Both analytical and numerical studies of black strings in 5D, spherically symmetric spacetime metrics have shown such strings to be unstable \cite{Scheel2014}. This is a significant criticism of Wesson's 5D non-compactified Kaluza-Klein theory (sometimes called the Space-Time-Matter or STM theory) that depends on an additional spatial dimension to generate matter \cite{Overduin:1997}. The equations \ref{eqn:pi2}, on the other hand, do not suffer from the same failure because they represent a chaotic flow around the classical 4D Einstein theory with imaginary time rather than an extension of Einstein's theory to 5D. This causes the extra spatial dimension to have a temporal rather than spatial nature, where we define temporal to mean both that there is a flow in that dimension and that the dimension behaves like time in the wave linearization of the equations.

We do not undertake to do an exhaustive review of stability results and whether they apply to these equations, which is outside the scope of this paper. Rather, this section investigates static spherically symmetric solutions with a restricted goal of establishing the value of $\epsilon$. We will show with a simple Regge-Wheeler, weak perturbation analysis that, while these solutions appear stable when $\epsilon=-1$, when $\epsilon=1$, they are unstable, indicating that $x_4$ is likely spacelike.

Let the metric have the following isotropic form (following the conventions in \cite{Wald2010}),
\[
dS^2 = -e^fdt^2 + e^h[dr^2 + r^2(d\theta^2 +\sin^2{\theta} d\phi^2)] - \epsilon d\tau^2
\] with $f$ and $h$ functions of $r$ and $\tau$. The Wick rotation back to real time modifies all the instances of $\sqrt{\gamma}\rightarrow\sqrt{-\gamma}$. We use the dotted notation $\dot{f}=\partial f/\partial\tau$ for $\tau$ derivatives and primes $f'=\partial f/\partial r$ for derivatives in $r$.

The components of $\pi^{\alpha\beta}$ are 
\begin{eqnarray}
\pi^{00} &=& 3e^{(3h-f)/2}r^2|\sin\theta| \dot{h},\\
\pi^{11} &=& -e^{(f+h)/2}r^2|\sin\theta|(\dot{f}+2\dot{h}),\\
\pi^{22} &=& -e^{(f+h)/2}|\sin\theta|(\dot{f} + 2\dot{h}),\\
\pi^{33} &=& -(e^{(f+h)/2}/|\sin\theta|)(\dot{f} + 2\dot{h}).
\end{eqnarray}

The components of $U^{\alpha\beta}$ are (lowering one index),
\begin{eqnarray}
	U^{0}{}_0 &=& -\frac{3}{4}\Bigg[-\gamma \dot{f}^2-2\left(1+5\gamma+6\sqrt{-\gamma}\right)\dot{f}\dot{h}+\left(6-13\gamma-12\sqrt{-\gamma}\right)\dot{h}^2+4\ddot{h}\Bigg],\\
	\label{eqn:U0}
	U^{1}{}_1 &=& \frac{1}{4}\Bigg[\left(-2+3\gamma + 4\sqrt{-\gamma}\right)\dot{f}^2+2\left(-3+15\gamma + 14\sqrt{-\gamma}\right)\dot{f}\dot{h}+\nonumber\\ &
	 &\left(-4+39\gamma+40\sqrt{-\gamma}\right)\dot{h}^2-4(\ddot{f}+2\ddot{h})\Bigg],\\
	 \label{eqn:u1}
	U^{2}{}_2 &=& U^3{}_3 = U^1{}_1 
\end{eqnarray} where $-\gamma=e^{f+3h}r^4\sin^2\theta$.

The Schwarzschild solution clearly solves the equations for $U^{\alpha\beta}=0$ which is true if $\partial_4\gamma_{\alpha\beta}=0$.  If $\partial_4\gamma_{\alpha\beta}\in O(\delta\gamma_{\alpha\beta})$ for some small $\delta\ll 1$, then, to first order, $U^{\alpha\beta} \approx (\gamma^{\alpha\mu}\gamma^{\beta\nu} - \gamma^{\alpha\beta}\gamma^{\mu\nu})\partial_4^2\gamma_{\mu\nu}$. Given that $T^{\alpha\beta}=0$, we can see that since $R^{\alpha\beta}=0$ for the Schwarzschild solution, the solution has no acceleration. We can ask, however, whether the Schwarzschild solution is stable for a given choice of $\epsilon$. That is, given that $\partial_4$ is small, does it remain small if we perturb the metric?

It turns out that, in this approximate form, the $x_4$ dimension takes on the role of time in the Regge-Wheeler equations but only if $x_4$ is spacelike ($\epsilon=-1$). Otherwise, it behaves like another spatial dimension. In addition, the analysis must be done under Wick rotation in order to satisfy the equivalence between statistical equilibrium theory and quantum theory. In the following, I look at first order stability of a spherically symmetric Schwarzschild solution (sometimes just called a wormhole) \cite{Regge1957}\cite{Zerilli1970}\cite{Fiziev2007}.

The extension of Regge-Wheeler ``odd'' parity, which I will refer to as ``magnetic'', spherical harmonics is straightforward. Assume that we are interested in spherical harmonics with coefficients as functions of $x_0=T$, $x_4=W$, and $x_1=r$, $A(W,T,r)Y_L^M(\theta,\phi)$. Also assume that parity $L$ and $M$ ($=0$) are constants of motion. The dependence on $T$ and $W$ looks like $e^{-ik_4 W - ik_0 T}$ for the perturbation  $h_{\mu\nu}$ around the static Schwarzschild background metric, $\gamma_{\mu\nu}$. Take the Wick rotation, $T\rightarrow i{\mathcal T}$ and $k_0\rightarrow ik_0$ and the dependence is $e^{-ik_4 W + ik_0 {\mathcal T}}$.

Going through some calculations, such as a gauge transformation to the Regge-Wheeler gauge, using the equation \ref{eqn:pi2} for small $\partial_4$, and we find that we end up with a second order wave equation similar to the Regge-Wheeler equation,
\begin{equation}
\partial^2 Q/\partial r^{*2} - [\epsilon k_4^{2} + k_0^2 + V_{eff}(r)] Q = 0,
\label{eqn:rw}
\end{equation} where $V_{eff} = (1-2m/r)[L(L+1)/r^2 - 6m/r^3]$ and $r^* = r + 2m\ln(r/2m -1)$ is Wheeler's {\em tortoise} coordinate.  Note that the 5th dimension is in the role of time and time has taken on a spatial dimension along with $r$. The function $Q=(1-2m/r)h_1/r$, where $h_1$ is one of two a scalar perturbation functions in the RW gauge for magnetic parity. (Note $m$ is the mass of the central object in units such that $G=c=1$.) 

Thus, it follows that,
\[
(\partial^2/\partial r^{*2} + \partial^2/\partial {\mathcal T}^2 + \epsilon \partial^2/\partial W^2)Qe^{-ik_4 W + ik_0 {\mathcal T}} = V_{eff}(r).
\] 

For ``even'', i.e. electric, perturbations the equation is identical but with a different effective potential \cite{Zerilli1970} or with a different definition of $Q$ \cite{Chandrasekhar1975} exactly as in standard Schwarzschild perturbation theory.

Near the Schwarzschild radius $r^* = 2m$ and far away $r^* \gg 2m$, for $\epsilon=-1$ perturbations behave like waves propagating in a two dimensional space $r^{*2} + {\mathcal T}^2 \leq W^2 + \mathtt{Constant}$. Since the wave equation is two dimensional rather than one dimensional, it implies that, rather than a standing wave propagating along a radial line in time, the perturbations propagate (in imaginary time) along a surface or membrane made up of the radial dimension and imaginary time into the fifth dimension.

Solving \ref{eqn:rw} as plane waves going into the wormhole and out of it, asymptotically as $r^*\rightarrow 2m$ and $r^*\rightarrow\infty$,
\begin{eqnarray}
	Q &\approx& c_1 e^{i\delta}(r^*/2m - 1)^{2i\sqrt{-\epsilon k_4^2-k_0^2}m} + c_1 e^{-i\delta}(r^*/2m - 1)^{-2i\sqrt{-\epsilon k_4^2-k_0^2}m},\quad r^*\rightarrow 2m,\\
	Q &\approx& c_2\sin (r^*\sqrt{-\epsilon k_4^2-k_0^2} + \eta),\quad r^*\rightarrow\infty,
\end{eqnarray} where $\delta$ is a small perturbation. For $\epsilon=-1$, instabilities, with exploding amplitude, occur when $k_4^2 < k_0^2$ exactly as they do when $k^2<0$ in the standard RW theory. Given the regularity of $Q(r^*)$ at $r^*=2m$ and $r^*=\infty$, the equation \ref{eqn:rw} is a self-adjoint eigenvalue problem for $k_4^2 - k_0^2$, meaning that $k_4^2 - k_0^2$ is real as well. 

For $\epsilon=1$, we have in \ref{eqn:rw} a Poisson equation in which perturbations diffuse into a three dimensional space of $r,W,{\mathcal T}$ as in heat diffusion with no wavefront. As in the case of imaginary $k$ in standard RW theory, this results in instability of the solution. Because solutions must be statistically asymptotically stable in $x_4$ in order to reach statistical equilibrium, the stability of the Schwarzschild  solution suggests that $x_4$ is spacelike. Thus, we can rule out $\epsilon =1$ completely.

For $V_{eff}=0$ which occurs in the two asymptotic regimes, we have the dispersion relation $k_4^2=k_0^2+k_1^2$. Thus, $k_4^2 \geq k_0^2$, and we conclude that, for spacelike $x_4$, the Schwarzschild solution is stable for weak perturbations.

The above analysis of stability can be extended to more spherically symmetric regimes (such as collapsed stars and other spherical bodies), and to include stronger and more dynamical perturbations as well as looking at intermediate regimes where $V_{eff}$ is not negligible. Even this basic analysis, however, eliminates the possibility of a timelike fifth dimension.

\subsection{Isotropic, Homogeneous Solution}
By making $f$ and $h$ functions of time $t$ rather than $r$, the isotropic equations \ref{eqn:pi2} and \ref{eqn:U0} and \ref{eqn:u1} become homogeneous in a spatially flat topology $k=0$ in the FLRW cosmology. If we assume that $U^{\alpha\beta}\neq 0$, then it behaves, on a cosmological scale, like matter. Assume that higher dimensional matter (which we assume is baryonic) and matter that is induced by geometry are each perfect fluids $T^{\alpha}{}_\beta = (\rho_b + p_b)u^\alpha u_\beta - \delta^\alpha{}_\beta p_b$, $\epsilon U^\alpha{}_\beta = (\rho_g + p_g)u^\alpha u_\beta - \delta^\alpha{}_\beta p_g$.

We know that $U^0{}_0 + U^1{}_1 - U^2{}_2 = \epsilon\rho_g$ and $\epsilon p_g=-U^2{}_2$ \cite{Overduin:1997}. Thus,
\begin{equation}
\rho_g = -\epsilon\frac{3}{4}\Bigg[-\gamma \dot{f}^2-2\left(1+5\gamma+6\sqrt{-\gamma}\right)\dot{f}\dot{h}+\left(6-13\gamma-12\sqrt{-\gamma}\right)\dot{h}^2+4\ddot{h}\Bigg]
\label{eqn:rhog}
\end{equation} and
\begin{eqnarray}
p_g & = & -\epsilon\frac{1}{4}\Bigg[\left(-2+3\gamma + 4\sqrt{-\gamma}\right)\dot{f}^2+2\left(-3+15\gamma + 14\sqrt{-\gamma}\right)\dot{f}\dot{h}+\nonumber\\ &
&\left(-4+39\gamma+40\sqrt{-\gamma}\right)\dot{h}^2-4(\ddot{f}+2\ddot{h})\Bigg].
\label{eqn:pg}
\end{eqnarray} 

Following \cite{Overduin:1997}, these can be split into four components $\rho_g = \rho_r + \rho_d + \rho_v + \rho_s$ and $p_g=p_r+p_d+p_v+p_s$. These correspond to radiation, $p_r=\rho/3$, dust, $p_d=0$, vacuum, $p_v=-\rho_v$, and stiff matter, $p_s=\rho_s$. For $\dot{f}\neq 0$ or $\ddot{f}\neq 0$, violations of the strong and weak energy conditions with zero density and non-zero pressure. While baryonic matter is not known to violate these conditions, it is not unphysical for extrinsic curvature to have negative energy. Nevertheless, such matter has not been observed.

If we assume $\dot{f}=0$, meaning $f$ is constant in $\tau$ by scaling the time coordinate so that $\gamma_{00}=-1$ as in standard FLRW theory, then we have only standard matter-like terms. With this assumption we match the terms in the equations \ref{eqn:rhog} and \ref{eqn:pg} and find that $\rho_s=\rho_r=0$, and we only have dust like matter and vacuum like matter:
\begin{eqnarray}
	\rho_d & = & -\epsilon\ddot{h} - \epsilon \left(\frac{7}{2} + \sqrt{-\gamma}\right)\dot{h}^2,	\label{eqn:rhod}\\
	\rho_v & = & -2\epsilon\ddot{h} - \epsilon \left(1 + \frac{39}{4}(-\gamma) - 10\sqrt{-\gamma}\right)\dot{h}^2.
	\label{eqn:rhov}
\end{eqnarray} Going to a non-flat spatial metric with $k=+1,-1$ in the FLRW model introduces factors of $(1-kr^2)$ into the instances of $\gamma$ and the vacuum density becomes,
\[\rho_v = -2\epsilon\ddot{h} - \epsilon \left(1 + \frac{39}{4(1-kr^2)}(-\gamma) - 10\sqrt{-\gamma}\right)\dot{h}^2.
\]

The equations \ref{eqn:rhod} and \ref{eqn:rhov} suggest that $\tau$ dependence in $h$ introduces both a cold dark matter and a cosmological dark vacuum energy term into the equations. If $\epsilon=-1$ as indicated above, then positive spatial growth $\dot{h}>0$ that is neutral ($\ddot{h}=0$) or accelerating ($\ddot{h}>0$) in the $\tau$ dimension creates positive energy density in both of these terms.

Liu and Wesson have investigated dynamical dark matter and energy in a Kaluza-Klein theory to show a Big Bounce model extension of the FLRW model \cite{Liu2001}. This model is similar but not directly applicable because they rely on the 5D vacuum Einstein equations. Our Hamiltonian quantization with higher dimensional matter has baryonic matter as an additional degree of freedom. Nevertheless, making certain assumptions about the relationship between $\tau$ and time in the metric such as either separation of variables or a wave-like form with a single parameter $\tau-Ct$ (as in \cite{Liu2001}) as well as imposing a condition on radiation versus dust such as $p=\eta \rho$ with $\eta=1/3$ all radiation and $\eta=0$ all dust can properly constrain the equations to admit a solution. A detailed investigation of these models and their solutions, however, is beyond the scope of this paper.

\section{Tests}
I have shown that the predictions of the theory match those of both general relativity and standard quantization when extrinsic curvature at scales above some small scale, e.g., the Planck scale, are negligible.

In the case where those derivatives are not negligible, the fully coupled equations of matter and gravity predict dust-like and vacuum-like terms that, in four dimensions, appear to be ``dark'' matter and energy. These terms arise from  $U^{\alpha\beta}$ on the right hand side of \ref{eqn:pi2} and contain extrinsic curvature. These must be dark because of the way gravity couples to electromagnetism, causing it to curve or shift its frequency but not itself emitting any light.

This extrinsic curvature is given by the coupling of baryonic matter in five dimensions to the gravitational metric given by the full equations \ref{eqn:pbra}. Thus, another prediction of the theory is that these dark matter and energy terms are tied to baryonic matter. Thus, they are tightly coupled. The extrinsic curvature's quadratic terms also indicate strong self-interaction at small scales, similar to Strongly interacting massive particle (SIMP) models for dark matter \cite{Wandelt2001}.

This prediction potentially addresses two problems. The first is the coupling of dark matter to baryonic in observations of the rotational curves and galaxies. In models of cold dark matter, such as dark matter halos, it is difficult to explain why dark matter appears to be so closely tied to baryonic matter. In this theory, the dark matter is simply the result of the baryonic matter's behavior in the fifth dimension.

The second problem is the coincidence problem where the proportion of dark energy in the universe is about 70\% of the total mass. This is suspicious because, in the current epoch, it is not considerably larger or smaller than the proportion of baryonic matter at about 5\%. If dark energy drives from extrinsic curvature induced by baryonic matter, however, its density must be connected to the density of baryonic which would provide an explanation for why it is similar.

Future work would include predicting probable densities, configurations, and interactions of these matter terms under reasonable assumptions.

\section{Quantum Interpretation}
The stochastic quantization framework provides an interpretation of path integrals, which are simply sums over configuration space, as dynamical behavior. Since path integrals, themselves, provide the underlying mathematics for the sum-over-histories interpretation of quantum measurement probabilities, the dynamical evolution undergirds that with an evolution mechanism that indicates exactly how human beings interact with the configuration space as observers. I call this the ``dynamic histories'' interpretation of quantum mechanics.

Within this framework, it is simple to interpret quantum measurements and resolve paradoxes because it provides both the source of the sample space, which is the fifth dimension, for those measurements as well as the complex interaction and correlation between different samples that occurs as a consequence of evolution in the fifth dimension. In this sense, the five dimensional theory resolves all quantum paradoxes in the same way that Everett's Many Worlds Interpretation does \cite{Dewitt2015}, but with only a single world that is evolving in an additional dimension. Unlike the MWI, however, evolution in a fifth dimension provides a clear mechanism for the world ``splitting'' that is not dependent on a measurement being made. It happens whether observations occur or not. Nor do infinite ``timelines'' continue on independent of one another. Rather, there is only one timeline which may not be fixed.

In addition, unlike MWI, the dynamic histories interpretation does not necessarily allow for all possible worlds. As in ordinary statistical mechanics, the number of configurations, e.g., a box of gas, goes through in a fixed amount of time is sufficient to match the predictions of a statistical ensemble for pressure and temperature but has vastly fewer actual configurations. Statistical physics makes the assumption, rather, that the amount of time over which a measurement of the gas is made $T$ is much greater than the amount of time $\delta$ between state transitions. Thus, $T$ is effectively infinite. The same is true in this dynamic histories interpretation. The number of separate configurations the 4D universe experiences may be considerably smaller than the total number of possible ones, yet, the transition between quantum states in that dimension is sufficient to match predictions that take all possible configurations into account (the path integral).

Since the resolution to quantum paradoxes is virtually identical to that of the MWI, a long exposition is not required on each one. One must simply exchange ``world'' for 4D slice. A few comments on well-known paradoxes follows:

The key point is that all quantum measurements derive from a sample space that is spread over the $\tau$ dimension. Since information cannot persist from one 4D slice to another since they are in statistical equilibrium, all measurements necessarily come from a single slice.

Schr\"odinger cat states \cite{Schrodinger1935}, for example, contain a superposition of two contradictory states, e.g., ``alive'' and ``dead''. Such states cannot both be contained in the same 4D slice but must exist at different sets of $\tau$, $A = \{\tau | \mathrm{cat\,is\,alive} \}$ and $D = \{\tau | \mathrm{cat\,is\,dead}\}$. A superposition of states is a statement of uncertainty about whether the $\tau'$ the observer exists when the measurement is observed is in $A$ or $D$. Hence, such states are not paradoxical but simply a statement of Bayesian uncertainty over a fifth dimension.

Einstein-Podolsky-Rosen (EPR) style paradoxes \cite{Einstein1935} are a consequence of uncertainty in $\tau$ as well. Measurements between different observers are correlated with one another because they exist at the same 4D slice in $\tau$ when those measurements are compared.

The evolution of quantum fields in the $\tau$ dimension also creates observations like the wave-particle duality, as in the double-slit experiment, since, while we have particles worldlines at each individual $\tau$, they evolve in $\tau$ as sheets to form, in real time, wavelike probability distributions. These probability distributions are a clear feature of the path integral/sum over histories interpretation of Feynman \cite{Feynman1948}, but, when reinterpreted as deterministic evolution, a clear mechanism for it as well as the place of the observer in it (at one point in $\tau$) appears.

\section{Conclusion}
In this paper, I have presented a novel quantization mechanism which features chaotic general relativity, Parisi-Wu stochastic quantization, and deterministic chaotic systems and their equivalence to stochastic systems. I have demonstrated with a scaling analysis that general relativity in harmonic coordinate gauge becomes highly nonlinear at small scales. I propose that this is sufficient to suggest that the universe has at least one real additional dimension and that quantum physics can be interpreted based on equilibrium flow in imaginary time of the 4D universe. Human beings are not aware of this dimension in general because we flow through it like time (unable to move around in it at will), but, unlike time where entropy is always increasing, reality in this dimension is at maximal entropy already. Information, hence, cannot grow or shrink except at the smallest scales where we typically see quantum effects where fluctuations may occur. Thus we cannot record or measure our flow in this dimension, but only observe the consequences of it in quantum measurements and their apparent paradoxes. Tests of the theory at large scales include observations of dark matter and energy which are predictions of the theory. Future work will include detailed analyses of both these predictions in terms of their density.

\bibliography{q5d}

\end{document}